\documentclass[journal,draftcls,onecolumn,12pt,twoside]{IEEEtran}
\usepackage{graphicx}
\usepackage{subfigure}
\usepackage{balance}
\usepackage{colortbl}
\usepackage{multicol}
\usepackage{multirow}
\usepackage{tabularx}
\usepackage{hyperref}
\usepackage{colordvi}
\usepackage{color}
\usepackage{amsmath, amsfonts, amssymb} 
\usepackage[none]{hyphenat}
\usepackage[noadjust]{cite}
\usepackage{arydshln}
\usepackage{longtable}
\usepackage{url} 
\usepackage{float}
\usepackage{cancel}
\usepackage{esint}

\definecolor{orange}{rgb}{1,0.5,0}

\begin{document}

\title{Throughput Analysis for Wavelet OFDM in Broadband Power Line Communications}

\author{Freddy~A.~Pinto{--}Benel, ~\IEEEmembership{Student Member,~IEEE,}
        Manuel~Blanco{--}Velasco,~\IEEEmembership{Senior Member,~IEEE,}
        Fernando~Cruz{--}Rold{\'a}n,~\IEEEmembership{Senior Member,~IEEE.}
\thanks{This work was partially supported by the Spanish Ministry of Economy and Competitiveness through project Research Grants TEC2015-64835-C3-1-R.}
\thanks{F.~A.~Pinto{--}Benel, and M.~Blanco{--}Velasco, F. Cruz-Rold{\'a}n are with the Department of Teor{\'i}a de la Se{\~n}al y Comunicaciones,         Escuela Polit{\'e}cnica Superior de la Universidad de Alcal{\'a}, 28871 Alcal{\'a} de Henares (Madrid), SPAIN (phone: + 34 91885 66 93; fax: + 34 91 885 66 99; e-mail: freddy.pinto@uah.es, manuel.blanco@uah.es, fernando.cruz@uah.es).}
}


\maketitle

\begin{abstract}
Windowed orthogonal frequency-division multiplexing (OFDM) and wavelet OFDM have been proposed as medium access techniques for broadband communications  over the power line network by the standard IEEE 1901. Windowed OFDM has been extensively researched and employed in different fields of communication, while wavelet OFDM, which has been recently recommended for the first time in a standard, has received less attention. This work is aimed to show that wavelet OFDM, which basically is an Extended Lapped Transform-based multicarrier modulation (ELT-MCM), is a viable and attractive alternative for data transmission in hostile scenarios, such as in-home PLC. To this end, we obtain theoretical expressions for ELT-MCM of: 1) the useful signal power, 2) the inter-symbol interference (ISI) power, 3) the inter-carrier interference (ICI) power, and 4) the noise power at the receiver side. The system capacity and the achievable throughput are derived from these. This study includes several computer simulations that show that ELT-MCM is an efficient alternative to improve data rates in PLC networks. 
\end{abstract}


%

\IEEEpeerreviewmaketitle



\section{Introduction}

\IEEEPARstart{C}{hannel partitioning} methods have become part of several standards for broadband wireless and wireline communications. The idea behind these techniques is to convert a broadband frequency channel into a set of overlapping  and (nearly) orthogonal frequency-flat subchannels \cite{Cio16}.  Orthogonal Frequency Division Multiplexing (OFDM), with or without windowing, is the most widely recommended technique and employs cyclic prefix (CP) or zeros as redundant samples to carry out the channel partition \cite{Li06}. Recently, filter bank multicarrier (FBMC) systems have attracted a great deal of research attention, with the difference that the partitioning is performed by means of pulse shaping, with good properties in time and frequency, with no additional redundant samples \cite{Lin10b}. FBMC system is now proposed as an attractive alternative to OFDM as the modulation scheme of the fifth generation of wireless networks (5G) \cite{Far14,Bog15,Ger17,Asg16,Gre16,Kofi17} .

The popularity of power line communications (PLC) for smart grids, and for outdoors and in-home communication systems, has grown a great deal \cite{Rab15,Juw16,Tao12,Ma13,Tch14,Cru16,Pou16,Cano2016}. It can be a good solution for the `last mile' problem, since it provides broadband communication to isolated places where other communication systems are not in place, and for the `last inch' problem, since it implements indoor high-speed networks \cite{Majumber2004}. IEEE Std 1901 \cite{IEEE10} defines a standard for broadband over power line (BPL) devices via electric power lines. All classes of BPL devices are considered for the use of this standard, including BPL devices used for smart energy applications or applications in transportation platforms (vehicles). 

As medium access techniques for PLC, IEEE P1901 has promoted windowed OFDM and FBMC, referring to the latter as wavelet OFDM. This denomination is a misnomer because the recommended system is not based on wavelet, it is a multicarrier modulation (MCM) based on the Extended Lapped Transform (ELT) \cite{Mal92,Cru16}. Hereinafter, wavelet OFDM is also referred to as ELT-MCM.

It is well known that one important drawback of OFDM is its insertion of redundancy, which reduces the throughput and the system capacity. As an alternative, FBMC is  a viable and attractive solution for communications over the mains network, because it does not require any kind of redundancy. However, channel equalization represents one of the most challenging issues and plays a major role in the PLC receiver based on FBMC system, chiefly since the PLC communication channel varies in both frequency and time, and experiences deep notches \cite{Mohammadi2009}.

Since OFDM with and without windowing has been recommended in other standards, e.g., HomePlug AV \cite{HomePlug}, it has received widespread attention by researchers. In this respect, there have been previous studies of the capacity and throughput of OFDM-based systems. For instance, in \cite{Lin10} the discrete multi-tone (DMT) system capacity was analyzed. The performance of windowed OFDM systems was studied in \cite{Achaichia2011, Nhan2014}. Previously, some other works focused their attention on FBMC systems, such as in \cite{Lek06, Ume06, Izu07} in which the transmission filters are obtained using the conventional cosine-modulated filter bank \cite[eq. (8.1.38)]{Vai93}. Recent literature has proposed contributions with specific emphasis on OFDM/OQAM (FBMC/OQAM) \cite{Achaichia2011,Rey17, Cans16}. To the best of the authors' knowledge, the study of the capacity and the throughput for the system recommended by the IEEE 1901 is still an open problem. The main purpose of this paper is to derive theoretical expressions for the wavelet OFDM system (or ELT-MCM), also including an Adaptive Sine-modulated/Cosine-modulated filter bank Equalizer for Transmultiplexer (ASCET) \cite{Alh01, Cru16}. In order to highlight the main advantages possessed by ELT-MCM as a medium access technique for broadband PLC, a comparison of the throughputs for windowed OFDM and ELT-MCM is included, considering an in-home PLC scenario.

The rest of this paper is organized as follows. In Section  \ref{M-FBMC}, the ELT-MCM system is briefly presented. In Section \ref{TE} theoretical expressions for the power of the useful signal, of the inter-symbol interference (ISI), of the inter-carrier interference (ICI), and of the noise, all at the receiver side, are derived. In Section \ref{capacity}, we present the transmission capacity. Section \ref{example} contains some simulation results. Section \ref{conclusions} provides our conclusions.

\section{Filter Bank Multicarrier Transceiver} \label{M-FBMC}
\begin{figure*}
\centering{
{\includegraphics[width=0.85\columnwidth]{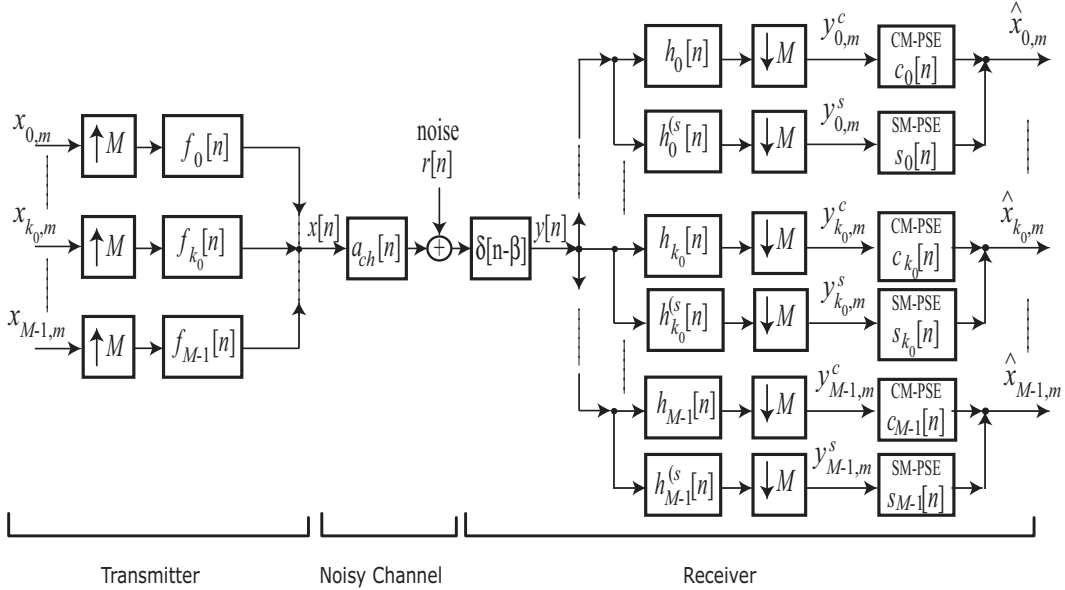} }
}
 \caption{General block diagram of an ELT-MCM with ASCET.} \label{tmux_ascet}
\end{figure*}

Fig. \ref{tmux_ascet} shows the block diagram of the filter bank multicarrier transceiver detailed in \cite{Cru16} and considered in the present paper. The baseband ELT-MCM physical layer recommended in \cite{IEEE10} is the following transmitting filter (for the $k$th subband, for $0\leq k\leq M-1$):  
\begin{equation} \label{fk_n}
f_k [n] = \sqrt {\frac{2}{M}} \cdot p[n] \cdot \cos \left[ { \left( { k + \frac{1}{2}} \right) \frac{\pi }{{M}}\left( {n + \frac{M+1}{2}} \right) } \right] \cdot \cos \left( \theta_k \right),
\end{equation}
where $0\leq n\leq N$, $M$ is the number of subbands, $p[n]$ is the prototype filter with length equal to $N+1=2 \kappa M$, and $\theta(k)$ is a phase constant equal to $0$ or $\pi$. Expression \eqref{fk_n}, excluding the term $\cos \left( \theta_k \right)$, is nothing but the ELT synthesis filters introduced by H. Malvar \cite{Mal92}. For the above transmitting bank, the reception system can be implemented as the time reflection of the transmission bank \cite{Cru16}: 
\begin{equation} \label{hk_n}
h_k [n]   =  \sqrt {\frac{2}{M}} \cdot p[n] \cdot \cos \left[ {\left( {k   +   \frac{1}{2}} \right)\frac{\pi }{M}} \cdot {   \left( {N - 1 - n + \frac{{M + 1}}{2}} \right)  } \right] \cdot \cos \left( \theta _k \right). 
\end{equation}
In order to compensate for the channel effects, the ASCET system can be used \cite{Vih06}. Its use for ELT-MCM was proposed in \cite{Cru16}, and the impulse responses of the receiving filters of the sine modulated filter bank (SMFB) are given by
\begin{equation} \label{hk_n_sm}
h_k^{s} [n]   =  \sqrt {\frac{2}{M}} \cdot p[n] \cdot \sin \left[ {\left( {k   +   \frac{1}{2}} \right)\frac{\pi }{M}} \cdot {   \left( {N - 1 - n + \frac{{M + 1}}{2}} \right)} \right] \cdot \cos \left( \theta _k \right). 
\end{equation}

With regard to $p\left[ n \right]$, different prototype filters are proposed in the standard \cite{IEEE10} for the cases of $M=512$, $1024$ and $2048$ subchannels, and for an overlapping factor $\kappa$ equal to $2$ or $3$. It is worth noting that the proposed prototype filters have even symmetry $\left( { p[N-1-n] = p[n]} \right)$. The standard does not provide expressions that allow designers to quickly obtain the corresponding coefficients, but in \cite{Cru16} it is shown that these prototype filters belong to a family of windows proposed by H. Malvar \cite{Mal92} which has the perfect reconstruction (PR) property in the context of filter banks. 

For more details of the transceiver configuration and for a quick and efficient way of implementing the whole system of Fig. \ref{tmux_ascet} by means of polyphase filters, we refer the reader to \cite{Cru16}.

\section{Analytical Expressions} \label{TE}
In Fig. \ref{tmux_ascet}, let us consider the discrete-time transmitted signal given by
\begin{equation}
x[n]= \sum_{k\in \mathbb{K}_{on}} \sum_{m\in \mathbb{Z}} x_{k,m}\cdot f_{k}\left[n-mM\right],
\label{eq:sn}
\end{equation}
where ${\mathbb{K}}_{on}  \subseteq \left\{ 0, \cdots, M-1 \right\} $ is the set of active subchannels defined by the tone mask \cite{IEEE10}, and the $x_{k,m}$ are the symbols in the subcarrier-time position $(k,m)$, assumed to be independent and identically distributed for every $k$ in ${\mathbb{K}}_{on}$. We assume that the PLC channel can be modeled as a time-invariant frequency selective channel:
\begin{equation}
a_{ch}[n]=\sum_{l=0}^{L-1}a_{l} \cdot \delta[n-l]\label{eq:hchn},
\end{equation}
where $L$ is the length of the channel. We also assume that there is no significant variation during a frame transmission. The received signal is the convolution of the transmitting signal with the channel impulse response plus an additive noise:
\begin{equation}
y[n]=\sum_{l=0}^{L-1}a_{l} \cdot x[n-l-\beta]+r[n-\beta],
\end{equation}
where $\beta$ is a delay included to obtain proper system operation. In this paper, we assume the noise to be additive white Gaussian noise with zero mean and variance $\sigma_r^2$. Lastly, the signal at the $i$-subcarrier output can be written as 
\begin{eqnarray}
y_{i,n}^{c}  &=& \sum\limits_{t = 0}^N {h_i \left[ t \right] \cdot y\left[ {nM - t} \right]} \nonumber \\
 &= &\sum\limits_{t = 0}^N {h_i \left[ t \right] } \cdot {\sum\limits_{l = 0}^{L - 1} {a_l  \cdot \sum\limits_{k\in \mathbb{K}_{on}} {\sum\limits_{m \in \mathbb{Z} }{x_{k,m} \cdot f_k \left[ {nM -t  - l -\beta- mM} \right]} } } }  \nonumber \\
&& + \sum\limits_{t = 0}^N {h_i \left[ t \right] \cdot r\left[ {nM -\beta - t} \right]} .
\end{eqnarray}
In the following subsections, we will obtain the signal-to-interference-plus-noise ratio (SINR) under the reasonable assumption that the number of subcarriers used is quite large, so that the interference on a given subcarrier is normally distributed \cite{Nhan2014}.
 

\subsection{Transmitting over a channel without noise} \label{without noise}

\noindent Noise apart, the output symbol at the position $\left(k_0,m_0\right)$ can be written as

\begin{eqnarray}
y_{k_{0},m_{0}}^{c}& = & \sum\limits_{k\in \mathbb{K}_{on}}  \sum_{m\in\mathbb{Z}}x_{k,m} \sum \limits_{l = 0}^{L - 1} {a_l } \cdot G_{k_0 ,m_0 }^c \left( {k,m,l} \right),
 \label{eq:yc_k0m0}
\end{eqnarray}
where 
\[
G_{k_0 ,m_0 }^c \left( {k,m,l} \right) = \sum\limits_{t = 0}^N {f_k } \left[ {m_0 M - t - l-\beta - mM} \right] \cdot h_{k_0 } [t].
\] 

The expression \eqref{eq:yc_k0m0} can be separated into the signal, $(k,m)=(k_0,m_0)$, and interferences, $(k,m) \neq (k_0,m_0)$. The first part gives rise to 
\begin{equation}\label{eq:k_m_0}
\Psi_{k_0, m_0}^c =  x_{k_{0},m_0}  \sum \limits_{l = 0}^{L - 1} {a_l } \cdot G_{k_0 ,m_0 }^c \left( {k_0,m_0,l} \right) = x_{k_{0},m_0} \cdot Q_{k_0 ,m_0 }^c \left( {k_0,m_0} \right),
\end{equation}
and the second one to
\begin{equation}\label{eq:Ic}
I_{k_0, m_0}^c = \sum \limits_{\scriptstyle {m\in\mathbb{Z}} \hfill \atop 
  \scriptstyle {m \neq m_0 }  \hfill} x_{k_{0},m}  \sum \limits_{l = 0}^{L - 1} {a_l } \cdot G_{k_0 ,m_0 }^c \left( {k_0,m,l} \right) = \sum \limits_{\scriptstyle {m\in\mathbb{Z}} \hfill \atop 
  \scriptstyle {m \neq m_0 }  \hfill} x_{k_{0},m} \cdot Q_{k_0 ,m_0 }^c \left( {k_0,m} \right),
\end{equation}
and
\begin{equation}\label{eq:Jc}
J_{k_0 ,m_0 }^c =  \sum \limits_{\scriptstyle {k\in \mathbb{K}_{on}} \hfill \atop 
  \scriptstyle {k \neq k_0 }  \hfill} \sum \limits_{ m \in \mathbb{Z}} {x_{k,m} }  \cdot \sum \limits_{l = 0}^{L - 1} {a_l } \cdot G_{k_0 ,m_0 }^c \left( {k,m,l} \right) = \sum \limits_{\scriptstyle \mathbb{K}_{on} \hfill \atop 
  \scriptstyle {k \neq k_0 }  \hfill}  \sum \limits_{ m \in \mathbb{Z}} {x_{k,m} }  \cdot Q_{k_0 ,m_0 }^c \left( {k,m} \right),
\end{equation}
where
\begin{equation*}
Q_{k_{0},m_0}^{c}\left(k,m \right) = \sum \limits_{l = 0}^{L - 1} {a_l } \cdot G_{k_0 ,m_0 }^c \left( {k,m,l} \right).
\end{equation*}
Therefore, the CMFB output symbol can be expressed as
\begin{equation}
y_{k_{0},m_{0}}^{c}=\Psi_{k_{0},m}^{c}+I_{k_{0},m_0}^{c}+J_{k_{0},m_{0}}^{c}.\label{eq:yc_final}
\end{equation} 
Following the same reasoning, the SMFB output symbol at the position $\left(k_0,m_0\right)$ is
\begin{equation}
y_{k_{0},m_{0}}^{s}=\Psi_{k_{0},m}^{s}+I_{k_{0},m_0}^{s}+J_{k_{0},m_{0}}^{s}.\label{eq:ys_final}
\end{equation} where
\begin{equation}\label{eq:k_m_0-1}
\Psi_{k_0, m_0}^s = x_{k_{0},m_0}  \sum \limits_{l = 0}^{L - 1} {a_l } \cdot G_{k_0 ,m_0 }^s \left( {k_0,m_0,l} \right) = x_{k_{0},m_0} \cdot Q_{k_0 ,m_0 }^s \left( {k_0,m_0} \right),
\end{equation}
\begin{equation}\label{eq:Is}
I_{k_0, m_0}^s = \sum \limits_{\scriptstyle m\in\mathbb{Z} \hfill \atop \scriptstyle {m \neq m_0 }  \hfill} x_{k_{0},m}  \sum \limits_{l = 0}^{L - 1} {a_l } \cdot G_{k_0 ,m_0 }^s \left( {k_0,m,l} \right) = \sum \limits_{\scriptstyle m\in\mathbb{Z} \hfill \atop  \scriptstyle {m \neq m_0 }  \hfill} x_{k_{0},m} \cdot Q_{k_0 ,m_0 }^s \left( {k_0,m} \right),
\end{equation}
\begin{equation}\label{eq:Js}
J_{k_0 ,m_0 }^s = \sum \limits_{\scriptstyle {k \in \mathbb{K}_{on}} \hfill \atop 
  \scriptstyle {k \neq k_0 }  \hfill}  \sum \limits_{ m \in \mathbb{Z}} {x_{k,m} }  \sum \limits_{l = 0}^{L - 1} {a_l } \cdot G_{k_0 ,m_0 }^s \left( {k,m,l} \right)  = \sum \limits_{\scriptstyle{k\in \mathbb{K}_{on}} \hfill \atop 
  \scriptstyle {k \neq k_0 }  \hfill}  \sum \limits_{ m \in \mathbb{Z}} {x_{k,m} }  \cdot  Q_{k_0 ,m_0 }^s \left( {k,m} \right).
  \end{equation}
where 
\[
G_{k_0 ,m_0 }^s \left( {k,m,l} \right) = \sum\limits_{t = 0}^N {f_k } \left[ {m_0 M - t - l-\beta - mM} \right] \cdot h_{k_0 }^s [t].
\]

\subsection{Noise Effects} \label{Noise}

The channel noise in an in-home PLC channel results from the contribution of different noises \cite{Zim02, Bar11}. Taking this into consideration, \eqref{eq:yc_final} and \eqref{eq:ys_final} can be rewritten as
\begin{equation}
y_{k_{0},m_{0}}^{c}=\Psi_{k_0 ,m_0 }^c+I_{k_{0},m_0}^{c} +J_{k_{0},m_{0}}^{c}+r_{k_{0},m_{0}}^{c},
\end{equation}
\begin{equation}
y_{k_{0},m_{0}}^{s}=\Psi_{k_0 ,m_0 }^s+I_{k_{0},m_0}^{s}+J_{k_{0},m_{0}}^{s}+r_{k_{0},m_{0}}^{s},
\end{equation}
where
\begin{equation}
r_{k_{0},m_{0}}^{c}=\sum_{t=0}^{N}{h}_{k_{0}}[t]\cdot r\left[m_{0}M-\beta-t\right],
\end{equation}
\begin{equation}
r_{k_{0},m_{0}}^{s}=\sum_{t=0}^{N}{h}_{k_{0}}^{(s}[t]\cdot r\left[m_{0}M-\beta-t\right].
\end{equation}


\subsection{The SINR of the 0-ASCET} \label{0-ASCET}
In the 0-ASCET \cite{Alh01,Vih06}, the cosine-modulated per-subcarrier equalizer (CM-PSE) and the sine-modulated (SM) PSE are memoryless FIR filters whose impulse responses are constants\footnote{Background material on the design of an ASCET for ELT-MCM can be found in \cite{Cru16}.}:
\[
c_k \left[ n \right] = c_k, \qquad s_k \left[ n \right] = s_k.
\]
Thus, the demodulated $\left(k_0,m_0\right)$th symbol in the absence of noise can be written as
\begin{eqnarray}
\hat{x}_{k_{0,}m_{0}}&=&y_{k_{0},m_{0}}^{c}\cdot c_{k_{0}}+y_{k_{0},m_{0}}^{s}\cdot s_{k_{0}} = \underbrace{\Psi_{k_{0},m_{0}}^{c}\cdot c_{k_{0}}+\psi_{k_{0},m_{0}}^{s}\cdot s_{k_{0}}}_{\Psi_{k_{0},m_{0}}^T} \nonumber\\
&&+\underbrace{I_{k_{0},m_{0}}^{c}\cdot c_{k_{0}}+I_{k_{0},m_{0}}^{s}\cdot s_{k_{0}}}_{I_{k_{0},m_{0}}^T} + \underbrace{J_{k_{0},m_{0}}^{c}\cdot c_{k_{0}}+J_{k_{0},m_{0}}^{s}\cdot s_{k_{0}},}_{J_{k_{0},m_{0}}^T}
\label{eq:x_detectado}
\end{eqnarray}
where $\Psi_{k_{0},m_{0}}^T$, $I_{k_{0},m_{0}}^T$ and $J_{k_{0},m_{0}}^T$ are the signal and the total interference parts at the $\left(k_0,m_0\right)$th symbol. 
The signal power can then be calculated as the second central moment:
\begin{equation}
P_{\Psi}\left(k_{0}\right) =  E\left[\left|\Psi_{k_{0},m_{0}}^T\right|^{2}\right]   =  \sigma_{x}^{2}  \Biggl| { Q_{k_{0},m_0}^{c} \left(k_0,m_0 \right) \cdot c_{k_{0}} }  { +Q_{k_{0},m_0}^{s}\left(k_0,m_0\right) \cdot s_{k_{0}} } \Biggr|^{2}.
\label{eq:Psi}
\end{equation}
Next, the intersymbol interference power is obtained as
\begin{equation}
P_{\text{ISI}}\left(k_{0}\right) = E\left[\left|I_{k_{0},m_{0}}^T\right|^{2}\right] = \sigma_{x}^{2}  \sum \limits_{\scriptstyle m\in\mathbb{Z} \hfill \atop  \scriptstyle {m \neq m_0 }  \hfill} \Biggl| { Q_{k_{0},m_0}^{c} \left(k_0,m \right) \cdot c_{k_{0}} }  { +Q_{k_{0},m_0}^{s}\left(k_0,m\right) \cdot s_{k_{0}} } \Biggr|^{2}.
\label{eq:Isi}
\end{equation}
On the other hand, the total intercarrier interference power can be obtained as
\begin{equation}
P_{\text{ICI}}\left(k_{0} \right) = E\left[\left|J_{k_{0},m_{0}}^T\right|^{2}\right] = \sigma_{x}^{2} { \sum \limits_{\scriptstyle \mathbb{K}_{on} \hfill \atop 
  \scriptstyle {k \neq k_0 }  \hfill}  \sum \limits_{ m \in \mathbb{Z}_0}  \Biggl| { Q_{k_{0},m_0}^{c} \left(k,m \right) \cdot c_{k_{0}} } { +Q_{k_{0},m_0}^{s}\left(k,m\right) \cdot s_{k_{0}} } \Biggr|^{2}}. 
 \label{Pint}
\end{equation}

The noise at the demodulated $\left(k_0,m_0\right)$th symbol can be expressed as
\begin{equation}
P_{r}\left(k_{0} \right)  =  E\left[\left| r_{k_{0},m_{0}}^{c}\cdot c_{k_{0}}+r_{k_{0},m_{0}}^{s}\cdot s_{k_{0}}\right|^2\right] =  \sigma_{r}^{2} \sum_{t=0}^{N}\left|c_{k_{0}}  \cdot {h}_{k_{0}} [t] + s_{k_{0}} \cdot {h}_{k_{0}}^{s}[t]\right|^{2} .
 \label{eq:Pr_final}
\end{equation}
Considering \eqref{eq:Psi}, \eqref{eq:Isi}, \eqref{Pint}, and \eqref{eq:Pr_final}, the SINR at the $k_0$th subcarrier is obtained as
\begin{equation}
\text{SINR}\left(k_{0}\right)=\frac{P_{\Psi}\left(k_{0}\right)}{P_{\text{ISI}}\left(k_{0}\right)+P_{\text{ICI}}\left(k_{0}\right)+P_{r}\left(k_{0}\right)}.
\label{eq:SINR}
\end{equation}
The theoretical expressions to obtain the SINR for a higher order ASCET are derived in Appendix \ref{L-ASCET}.

\section{System Capacity and Throughput} \label{capacity}
\subsection{Windowed OFDM PHY}

The theoretical throughput can be calculated by the following expression:

\begin{equation}
R= \Delta_{f} \sum_{k\in \mathbb{K}^{'}_{on}} \frac{M}{M+GI} \cdot C(k),\label{thro}
\end{equation} where $\Delta_{f}$ is the subcarrier spacing, $\mathbb{K}^{'}_{on}$ is the set of actives subcarriers, $GI$ is the length of the guard interval in samples, and $C(k)$ is the capacity of the $k$th subcarrier, which can be calculated by means of the following expression: 

\begin{equation}
C(k)=\log_{2}\left(1+\frac{\text{SINR}(k)}{\Gamma}\right). \label{Capacidad}
\end{equation}
$\Gamma$ is the SINR gap that, when $2^{2K}$-QAM constellation is used, is defined by

\begin{equation}
\Gamma \approx\frac{1}{3}\left[Q^{-1}\left(\frac{\text{SER}}{4}\right)^{2}\right],
\end{equation} in which $SER$ is the symbol error rate and $Q^{-1}$ is the inverse tail probability of the standard normal distribution \cite{Cio16, Lin10b}. On the other hand, the SINR of the windowed OFDM system using one-tap zero forcing (ZF) is derived in \cite{Achaichia2011}:
\begin{equation}
\text{SINR}(k)=\frac{\sigma^{2}_{x}(k)|H_k|^2}{\sigma^{2}_{n}(k)+P_{ICI+ISI}^{Wofdm}(k)},
\end{equation} 
where $\sigma^{2}_{x}$ is the variance of the transmitted symbols, $ \sigma^{2}_{n}$ denotes the variance of the PLC noise, $H_k$ is the channel frequency response coefficient, and $P_{ICI+ISI}^{Wofdm} (k)$ denotes the interference power, all the above on the $k$th subcarrier.

\subsection{Wavelet OFDM PHY}

As in the previous case, the ELT-MCM throughput can be obtained using \eqref{thro}. Since this method does not need a guard interval, and provided we assume that the interference power on a given subcarrier is normally distributed, the throughput formula yields

\begin{equation}
R=\Delta_f \sum_{k\in \mathbb{K}_{on}} \log_{2}\left(1+\frac{\text{SINR}(k)}{\Gamma}\right),\label{eq:capacidad}
\end{equation}
where the SINR is given by \eqref{eq:SINR} or \eqref{eq:SINR_Lascet}, and $\Gamma$ is also the SINR gap that models the practical modulation and coding scheme for the targeted SER. Since IEEE 1901 \cite{IEEE10} proposes the PAM as the primary mapping for the ELT-MCM PHY, and assuming that an $M$-QAM is like 2 $M$-PAM independent  modulations \cite{Cio16, Lin10b}, we have

\begin{equation}
\Gamma \approx \frac{1}{3}\left[Q^{-1}\left(\frac{\text{SER}}{2}\right)\right]^{2}.
\end{equation}

\section{Simulation Results} \label{example}

The theoretical throughput of the ELT-MCM and windowed OFDM system will be compared in this section considering an in-home PLC scenario. Following the same process as in \cite{Lin10, Achaichia2011}, ELT-MCM will be evaluated with a $0$-tap, $3$-tap and $5$-tap equalizer, while a zero-forcing equalizer will be used in windowed OFDM. 

\begin{figure}
\center{}
\includegraphics[width=0.6\columnwidth]{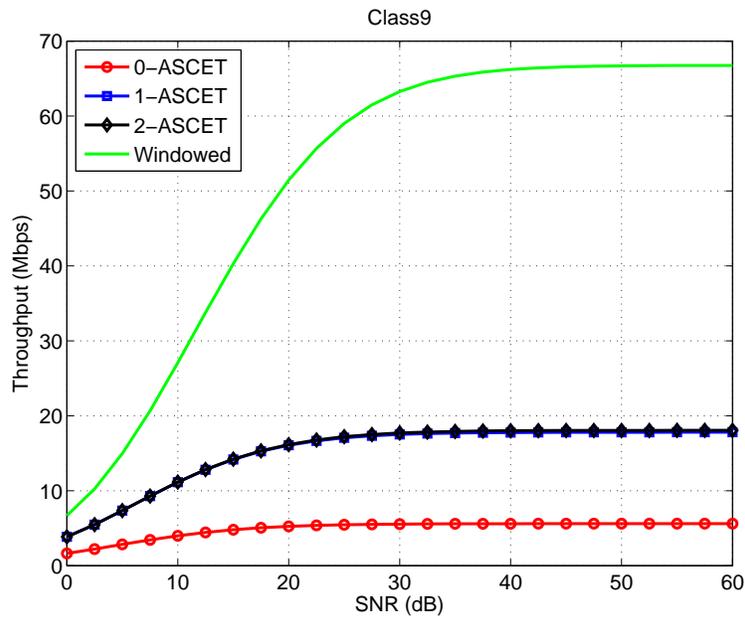}
\caption{Comparison of FFT OFDM PHY (windowed OFDM) and wavelet OFDM PHY (ELT-MCM) with 917 and 360 active subcarriers, respectively.\label{comp_512}}
\end{figure}

There are several approaches for modeling in-home PLC channels \cite{Zim02, Antoniali2011, Canete2011, Ton10}. In this paper, we focus our attention on the model proposed in \cite{Ton10}, which is a statistical approach that synthesizes different classes with a finite number of multi-path components. Our simulations consisted in averaging the outcomes of 100 transmissions through different impulse response realizations representative of class 1 (strong signal attenuation), class 5 (medium signal attenuation) and class 9 (little signal attenuation), which have been computed using the script available on-line in \cite{Ton10a}. In addition, we assume that the channel remains constant during each multicarrier symbol, and perfect channel knowledge is also assumed at the receiver. Furthermore, there is no kind of error-correcting-code. In particular, we assume the following conditions for each physical layer (PHY):

\begin{itemize}
\item Wavelet OFDM PHY:
Following the specifications recommended by IEEE P1901, the ELT-MCM system employs 512 subcarriers with 360 active subcarriers (used for data modulation) in the range from $1.8$ MHz to $28$ MHz and a frequency spacing ($\Delta_{f}$) of 61.03515625 kHz. For this transceiver, the performance is evaluated for the 0-ASCET, 1-ASCET and 2-ASCET \cite{Cru16}. 

\item FFT OFDM PHY:
The specifications of the windowed OFDM system are also based on \cite{IEEE10}. This system uses 4096 subcarriers with up to 1974 usable subcarriers in the range of $[1.8-50]$ MHz, and the carrier spacing is approximately 24.414 kHz. Support for carriers above 30 MHz is optional. Of the subcarriers below 30 MHz, 917 are active. In addition, the standard fixes a mandatory payload symbol guard interval (GI) equal to 556, 756 or 4712 samples. In order to ensure a fair comparison, approximately equal numbers of active subcarriers should be used for both transceivers. For this reason, we assume in our simulations $M=4096$ and 917 active subcarriers for the windowed OFDM, and furthermore, the GI is chosen to be 756 to provide good system performance and to not penalize this transceiver. A zero-forcing equalizer is assumed to compensate for the channel effects.
\end{itemize}
In the first simulation, the windowed OFDM throughput is compared with that of the wavelet OFDM, assuming for the latter $M=512$ and $360$ active subcarriers, and with class 9 in-home PLC channels. Fig. \ref{comp_512} depicts the theoretical throughput obtained under these conditions. Given that the number of active subcarriers in the windowed OFDM is almost three times larger than the number of the Wavelet OFDM usable subcarriers, a significant difference can be seen: the former outperforms the latter, almost tripling the achievable throughput at high SNR values. Moreover, since the PLC channel is one of the most hostile channels, the 0-ASCET is not enough to compensate for the channel distortion.

\begin{figure}
\center{}
\includegraphics[width=0.6\columnwidth]{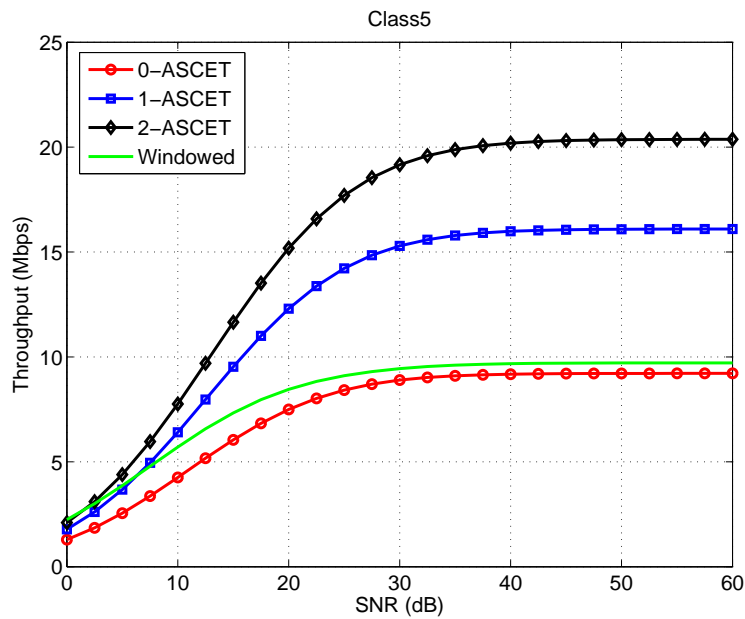}
\caption{Comparison of FFT OFDM PHY (windowed OFDM) with 917 and wavelet OFDM PHY (ELT-MCM) with 360 active subcarriers in the presence of class 5 PLC channels.\label{comp_512_class5}}
\end{figure}

In this second comparison, we investigate the performance of both methods under the same conditions as the first experiment, but in the presence of PLC channels of class 5. Fig. \ref{comp_512_class5} shows the resulting throughputs, and as can be seen, the highest values are achieved by the ELT-MCM system with 2-ASCET, even though the number of active subcarrier is one-third as much. These results are obtained when 1-ASCET or 2-ASCET are employed. Actually,  the ELT-MCM throughput associated with 2-ASCET is $179.5\%$ higher for SNR=20 dB. 


\begin{figure}
\center{}
\includegraphics[width=0.6\columnwidth]{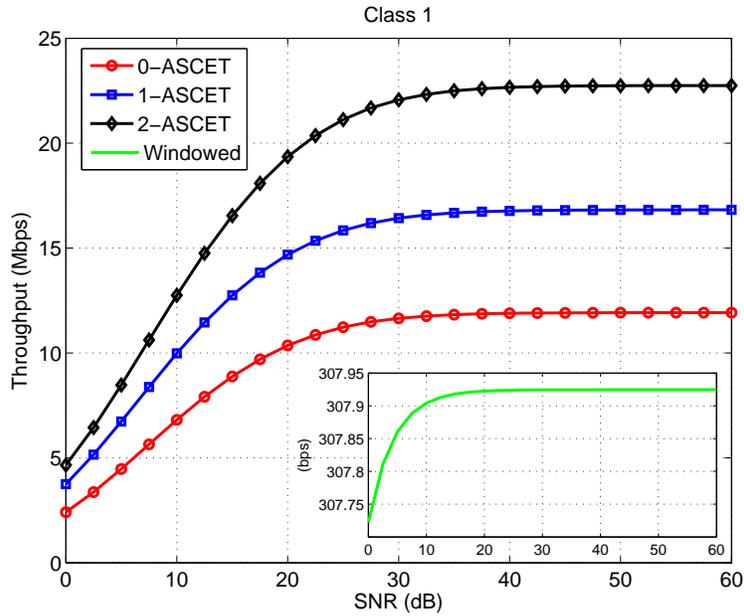}
\caption{Comparison of FFT OFDM PHY (windowed OFDM) with 917 and wavelet OFDM PHY (ELT-MCM) with 360 active subcarriers in the presence of class 1 PLC channels.\label{comp_512_class1}}
\end{figure}

As third scenario, both methods have been analyzed in the same conditions than in the first simulations but with class 1 PLC channels. As shown in Fig. \ref{comp_512_class1}, the throughput of windowed OFDM system is considerably reduced and it is outperformed by the ELT-MCM, which shows better results than those shown in the first simulation.


\begin{figure}
\center{}
\includegraphics[width=0.6\columnwidth]{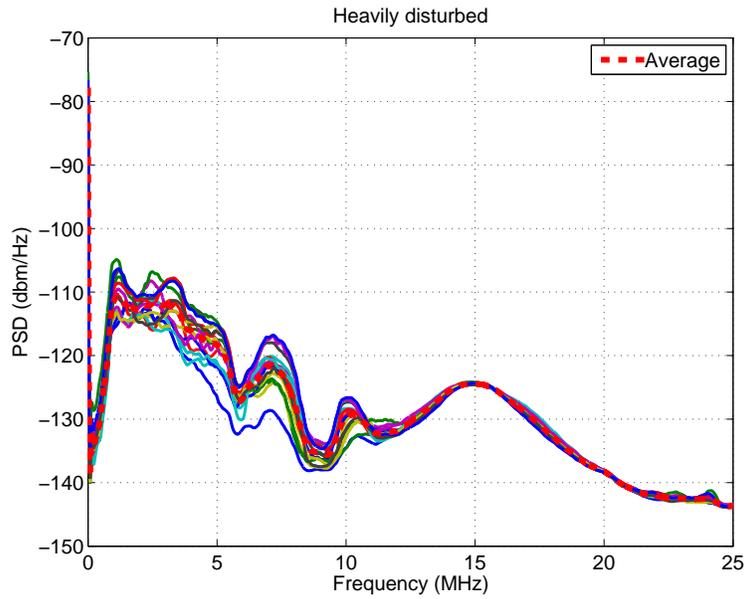}
\caption{Background noise PSD related to a heavily disturbed in-home channel.\label{comp_1024_class1}\label{BGN_fuerte}}
\end{figure}

In the following experiments, the AWGN has been replaced by a PLC background noise (BGN). The BGN has been modeled following \cite{Cortes2010} and it is the noise associated with a heavily disturbed channel. Its power spectral density is depicted in Fig. \ref{BGN_fuerte}. and it is referred to as strong BGN. 

\begin{figure}
\center{}
\includegraphics[width=0.6\columnwidth]{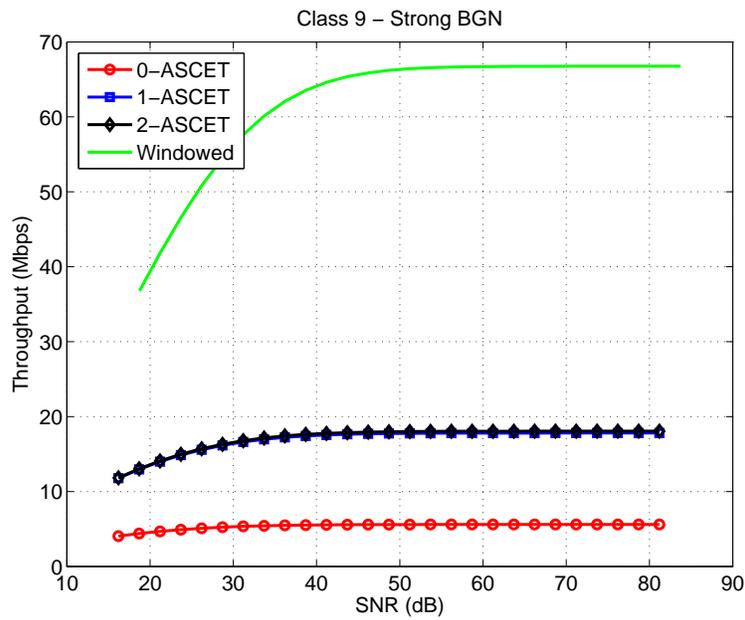}
\caption{Comparison of FFT OFDM PHY (windowed OFDM) and wavelet OFDM PHY (ELT-MCM) with 917 and 360 active subcarriers, respectively, in presence of strong BGN.\label{comp_512_class9_fuerte}}
\end{figure}

Next, the windowed OFDM throughput is compared with that of the ELT-MCM system, assuming for the latter $M=512$ and $360$ active subcarriers, and with class 9 in-home PLC channels. Fig. \ref{comp_512_class9_fuerte} depicts the theoretical throughput obtained under these conditions. Given that the number of  active subcarriers in the windowed OFDM is almost threefold larger than the number of the ELT-MCM usable subcarriers, a significant difference can be seen: the former outperforms the latter, almost tripling the achievable throughput at high SNR values. Moreover, the 0-ASCET is not enough to compensate for the channel distortion.

\begin{figure}
\center{}
\includegraphics[width=0.6\columnwidth]{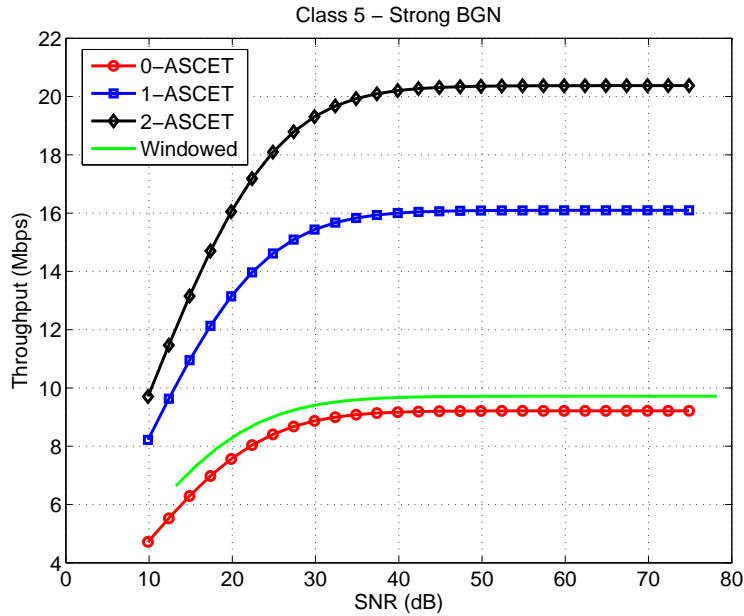}
\caption{Comparison of FFT OFDM PHY (windowed OFDM) with 917 and wavelet OFDM PHY (ELT-MCM) with 360 active subcarriers in the presence of class 5 PLC channels and strong BGN.\label{comp_512_class5_fuerte}}
\end{figure}

We investigate now the performance of both methods under the same conditions as the first experiment, but in the presence of PLC channels of class 5. Fig. \ref{comp_512_class5_fuerte} shows the resulting throughputs, and as can be seen, the highest values are achieved by the ELT-MCM system with 2-ASCET, even though the number of active subcarrier is one-third of that of the windowed OFDM. Actually,  the ELT-MCM throughput associated with 2-ASCET is $156\%$ higher for SNR=20 dB. 

\begin{figure}
\center{}
\includegraphics[width=0.6\columnwidth]{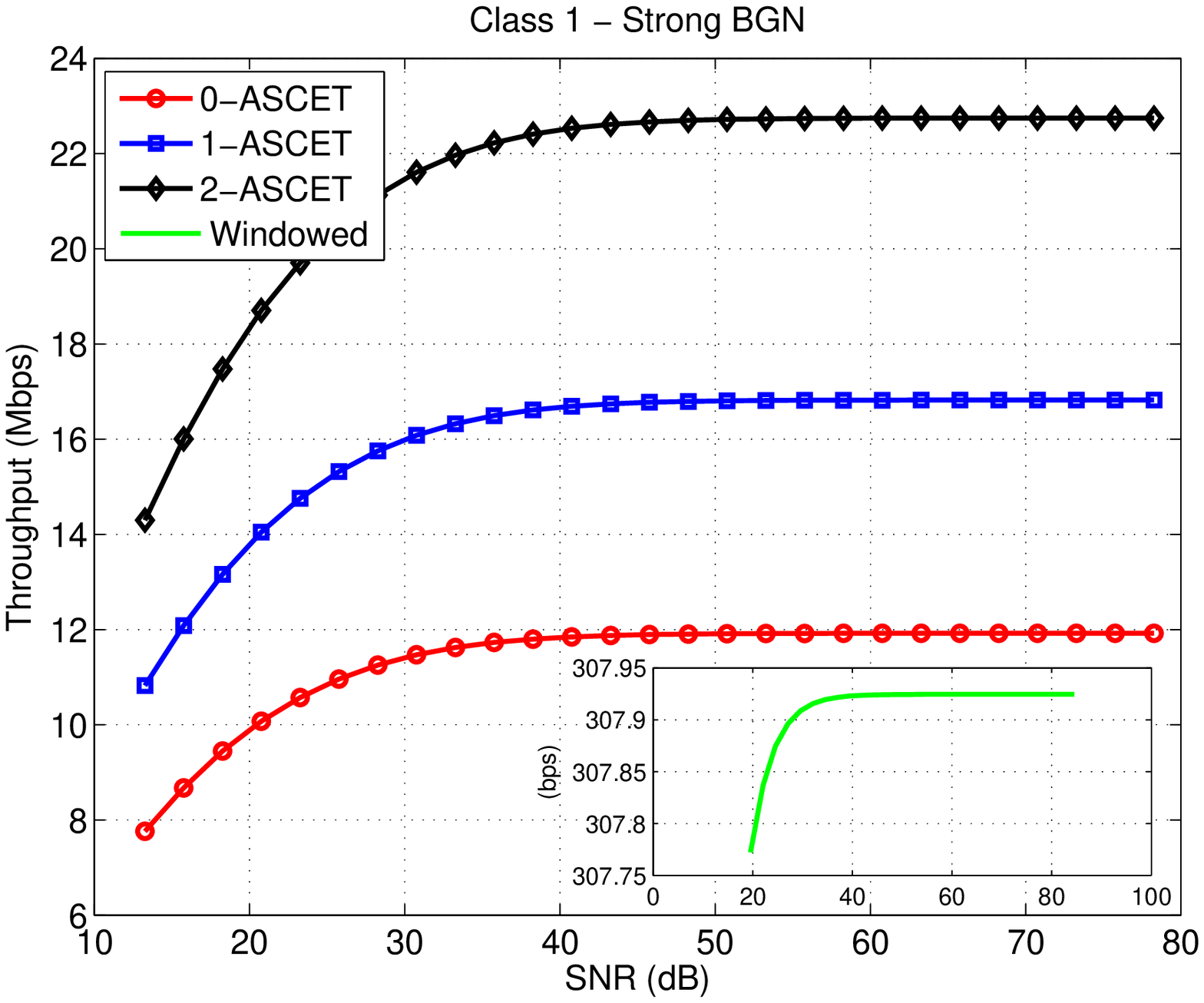}
\caption{Comparison of FFT OFDM PHY (windowed OFDM) with 917 and wavelet OFDM PHY (ELT-MCM) with 360 active subcarriers in the presence of class 1 PLC channel sand strong BGN.\label{comp_512_class1_fuerte}}
\end{figure}

Finally, both methods have been compared in the same conditions than in the first simulations but with class 1 PLC channels. As shown in Fig. \ref{comp_512_class1_fuerte}, the throughput of windowed OFDM system is considerably reduced and it is outperformed by the ELT-MCM, which shows better results than the results depicts in Fig. \ref{comp_512_class9_fuerte}

\section{Conclusion} \label{conclusions}
Wavelet OFDM is one of the medium access techniques adopted by the IEEE P1901 working group for broadband PLC. This paper has derived theoretical expressions for the different powers, corresponding to the useful signal, as well as to the intersymbol and the intercarrier interferences, and the noise, which are present at the receiver side for this kind of FBMC system. With these expressions, the capacity and the achievable throughput of wavelet OFDM for baseband communications  have been calculated. A performance comparison with windowed OFDM has also been provided, on the basis of simulations. In terms of achievable throughput, wavelet OFDM is a viable and attractive solution because it has outperformed the windowed OFDM in hostile PLC channels, even though using fewer active subcarriers.

\appendices
\section{The SINR of the $L_A$-ASCET}\label{L-ASCET}
The simplest equalizer with only one coefficient per subcarrier does not provide good performance in hostile channels. In order to improve its performance, the multiplications employed in 0-ASCET must be replaced by FIR filters, i.e.,
\[
c_k \left[ n \right] = \sum\limits_{\mu  =  - L_A }^{L_A } {c_{k,\mu }  \cdot \delta \left[ {n - \mu } \right]} ,
\]
\[
s_k \left[ n \right] = \sum\limits_{\mu  =  - L_A }^{L_A } {s_{k,\mu }  \cdot \delta \left[ {n - \mu } \right]} ,
\]
leading to a new equalizer of higher order, referred to as the $L_A$-order ASCET, and denoted by $L_A$-ASCET \cite{Alh01,Vih06,Cru16}. \\
In this case, the demodulated $\left(k_0,m_0\right)$th symbol, in the absence of noise when the system uses a $L_A$-ASCET, is
\begin{equation}
\hat{x}_{k_{0},m_{0}} = \sum_{\mu=-L_{A}}^{L_{A}}  \sum_{k\in \mathbb{K}_{on}} \sum_{m\in\mathbb{Z}}  x_{k,m}\cdot \Biggl(  Q_{k_0 ,m_0-\mu }^c \left( {k,m} \right)  \cdot  c_{k_{0},\mu } +   Q_{k_0 ,m_0-\mu }^s \left( {k,m}  \right) \cdot  s_{k_{0},\mu } \Biggr).
\end{equation}
As in the previous case, the equations can be split into two groups: the signal $(k,m)=(k_0,m_0)$ and the interferences $(k,m) \neq (k_0,m_0)$. For the first group,
\begin{equation}
\Psi_{k_{0},m_{0}}^T = \sum_{\mu=-L_{A}}^{L_{A}} x_{k_{0},m_0} \cdot \Biggl( Q_{k_{0},m_{0}-\mu}^{c}(k_0,m_0) \cdot  c_{k_{0},\mu}  + Q_{k_{0},m_{0}-\mu}^{s}(k_0,m_0 ) \cdot s_{k_{0}, \mu}  \Biggr),
\end{equation}
includes the useful symbol. Otherwise, the interference part can be defined as
\begin{equation}
I_{k_{0},m_{0}}^T = \sum_{\mu=-L_{A}}^{L_{A}} \sum \limits_{\scriptstyle m\in\mathbb{Z} \hfill \atop  \scriptstyle {m \neq m_0 }  \hfill} x_{k_{0},m} \cdot \Biggl( Q_{k_{0},m_{0}-\mu}^{c}(k_0,m) \cdot  c_{k_{0},\mu}   + Q_{k_{0},m_{0}-\mu}^{s}(k_0,m ) \cdot s_{k_{0}, \mu}  \Biggr),
\end{equation}
and
\begin{equation}\label{eq:eq_Lascet_kdis0_mdis0}
J_{k_{0},m_{0}}=\sum_{\mu=-L_{A}}^{L_{A}} \sum \limits_{\scriptstyle {k\in \mathbb{K}_{on}} \hfill \atop 
  \scriptstyle {k \neq k_0 }  \hfill} \sum \limits_{ m \in \mathbb{Z}}{x_{k,m} } \cdot \left(Q_{k_{0},m_{0}-\mu}^{c}\left(k,m \right)\cdot c_{k_{0},\mu} + Q_{k_{0},m_{0}-\mu}^{s}\left(k,m\right)\cdot s_{k_{0},\mu}\right).
\end{equation}
The useful signal power can be expressed as
\begin{equation}\label{eq:P_gamma}
P_{\Psi}\left(k_{0}\right) = \sigma_{x}^{2}  \sum_{\mu=-L_{A}}^{L_{A}} \Biggl| Q_{k_{0},m_{0}-\mu}^{c}\left(k_0,m_0 \right) \cdot c_{k_{0}, \mu} + \Biggl. Q_{k_{0},m_{0}-\mu}^{s}\left(k_0,m_0 \right) \cdot s_{k_{0}, \mu}  \Biggr|^{2}.
\end{equation}
On the other hand, the interference power can be obtained as
\begin{equation}\label{eq:P_ISI}
P_{\text{ISI}}\left(k_{0}\right) = \sigma_{x}^{2}  \sum_{\mu=-L_{A}}^{L_{A}} \sum \limits_{\scriptstyle m\in\mathbb{Z} \hfill \atop  \scriptstyle {m \neq m_0 }  \hfill} \Biggl| Q_{k_{0},m_{0}-\mu}^{c}\left(k_0,m \right) \cdot c_{k_{0}, \mu} + \Biggl. Q_{k_{0},m_{0}-\mu}^{s}\left(k_0,m \right) \cdot s_{k_{0}, \mu}  \Biggr|^{2},
\end{equation}
and
\begin{equation}\label{eq:PICI_Lascet}
P_{\text{ICI}}\left(k_{0}\right) = \sigma_{x}^{2}  \sum_{\mu=-L_{A}}^{L_{A}} \sum \limits_{\scriptstyle {k\in \mathbb{K}_{on}} \hfill \atop 
  \scriptstyle {k \neq k_0 }  \hfill} \sum \limits_{ m \in \mathbb{Z}} \Biggl| Q_{k_{0},m_{0}-\mu}^{c} \left( k,m\right)\cdot c_{k_{0}, \mu}   + Q_{k_{0},m_{0}-\mu}^{s}\left(k,m\right)\cdot s_{k_{0},\mu}  \Biggr|^{2}. 
\end{equation}

The noise power can be calculated as
\begin{equation}
P_{r}\left(k_{0}\right)=\sigma_{r}^{2}  \sum_{t=0}^{N+2L_{A}}\left|{h}_{k_{0},\mu} [t]  +  {h}_{k_{0},\mu}^{s}[t]\right|^{2} ,
\label{eq:Pr_Lascet}
\end{equation}
where
\[
{h}_{k_{0},\mu}\left[ n \right]= {h}_{k_{0}} \left[ n \right] *c_k \left[ n \right] ,
\]
\[
{h}_{k_{0},\mu}^{s}\left[ n \right]= {h}_{k_{0}}^{s}\left[ n \right] * s_k \left[ n \right] .
\]
As a result, \eqref{eq:P_gamma}, \eqref{eq:P_ISI}, \eqref{eq:PICI_Lascet} and \eqref{eq:Pr_Lascet} allow obtaining the SINR at the $k_0$th subcarrier:
\begin{equation}
\text{SINR}\left(k_{0}\right)=\frac{P_{\Psi}\left(k_{0}\right)}{P_{\text{ISI}}\left(k_{0}\right)+P_{\text{ICI}}\left(k_{0}\right)+P_{r}\left(k_{0}\right)}.\label{eq:SINR_Lascet}
\end{equation}





%

\bibliographystyle{ieeetr}
\balance
\bibliography{biblio_tcom}

\end{document}